\documentclass[aps,pra,showpacs,reprint]{revtex4-1}

\usepackage{graphicx}
\usepackage{amsthm,amsmath,amssymb,amsfonts}

\newcommand{\ket}[1]{| #1 \rangle}

\begin{document}

\begin{abstract}
We investigate the nature of the pulse sequence so that unwanted transitions in quantum systems can be inhibited optimally. For this purpose we show that the sequence of pulses proposed by Uhrig [Phys. Rev. Lett. \textbf{98}, 100504 (2007)] in the context of inhibition of environmental dephasing effects is optimal. We derive exact results for inhibiting the transitions and confirm the results numerically. We posit a very significant improvement by usage of the Uhrig sequence over an equidistant sequence in decoupling a quantum system from unwanted transitions. The physics of inhibition is the destructive interference between transition amplitudes before and after each pulse.
\end{abstract}  

\title{Optimized pulse sequences for suppressing unwanted transitions in quantum systems}
\author{C.A. Schroeder}
\email{schro87@ou.edu}
\affiliation{Homer L. Dodge Department of Physics and Astronomy, University of Oklahoma, 440 W. Brooks St. Norman, OK 73019}
\author{G.S. Agarwal}
\affiliation{Department of Physics, Oklahoma State University, Stillwater, Oklahoma 74078, USA}
\date{\today}
\pacs{03.67Pp, 82.56Jn, 42.50-p}
\maketitle


\section{Introduction}
The power of quantum information processing stems from the uniquely quantum phenomena of superposition and entanglement.  Quantum bits (qubits) are two-level systems with states we can associate with the $0$ and $1$ of classical bits, but the valuable superposition quality is additional infomation coded in phase information called coherence \cite{di}\cite{ni}. Coupling of a qubit to it's environment entangles the two sytems, which effects a partial measurement on the qubit and introduces phase errors, a process known as decoherence \cite{ge}\cite{pa}.  In NMR terminology, this corresponds to energy-conserving $T_2$-type decay \cite{sl}.  Qubits are also subject to non-energy conserving amplitude errors, or $T_1$-type decay.  Dynamical decoupling seeks to suppress errors with appropriate perturbations.  Application of $\pi$-pulses to a qubit-bath system -- dubbed quantum "bang-bang" decoupling -- effects a sort of time reversal in the system, which creates interference and partial cancellation of environmental effects \cite{pa}\cite{vi}\cite{da}.  Further work showed that random or concatenated pulse sequences could offer improvement over a sequence spaced equally in time \cite{kn}\cite{l1}\cite{l2}.  In this context Uhrig has shown that an optimal pulse sequence exists for decoupling a spin-1/2 particle from a bath of harmonic oscillators \cite{uh}\cite{uE}.  The improvement of this sequence for preserving coherence has also been verified experimentally \cite{du}\cite{bi}.  For a time interval $T$ and an $N$-pulse sequence, the Uhrig Dynamical Decoupling (UDD) sequence suppresses decoherence up to order $T^{N+1}$.  Yang then showed that UDD suppresses decoherence up to order $T^{N+1}$ for a more general qubit-bath Hamiltonian \cite{ya}.  Indeed, Yang shows that the UDD is optimal for suppressing $T_1$- or $T_2$-type decay-- corresponding to either phase or amplitude errors.  The ideas of Uhrig have been extended to protection of entanglement in two or more qubits \cite{Ag}.
	
	The present paper considers the problem of unwanted transitions (decay and excitation) between two atomic levels due to a perturbing field.  We examine a scheme which suppresses environmental effects by quasi-instantaneous phase changes on a two-level system \cite{ag}\cite{wa}.  By applying a sequence of ultrashort $2\pi$ pulses to one state of the two-level system through an auxilliary level, this state acquires a phase shift of $\pi$ with respect to the other (See Fig.1).  The phase shift leads to destructive interference between evolution terms in the time intervals between the pulses.  Note that unlike the traditional spin problems where $\pi$ pulses are applied between the two levels, here we apply $2\pi$ pulses between one of the levels and an auxilliary level.  This idea has been implemented in experiments in which fullerene qubits are decoupled from nuclear Rabi oscillations \cite{mo}.  The question that we investigate in this paper is: what is the optimum sequence of pulses to suppress unwanted transitions?  We will show that the UDD sequence is optimal for a two level transition in an external field in the semi-classical limit.

The plan of the paper is the following.  In Sec. II we prove the optimality of the Uhrig sequence using first-order perturbation theory in the case of large detuning.  In Sec. III we look at the exact case using the formalism of transfer matrices.  In Sec. IV we prove again the optimality of the UDD for zero detuning.  Sec. V reinforces these findings with numerical studies which suggest the optimality of the UDD across the all ranges of detunings.  Finally, Sec. VI discusses applications and conclusions.  In summary, we find that the UDD sequence offers a significant improvement over traditional equidistant pulse sequences for decoupling a two-level system from an unwanted transition.  
	
Furthermore, we posit that using the UDD in experiments would improve the decoupling by a factor $\sim \tfrac{1}{n} $ for resonant perturbations and by many orders of magnitude for increasingly detuned perturbations.  For example, in Figure 3 we reproduce numerically some data from an experiment by Morton, where the equidistant sequence was used \cite{mo}.  In addition, we have generated the data under the influence of the UDD sequence to illustrate graphically the improvement offered by such a sequence.

\begin{figure}
\centering 
\includegraphics[width=8.6cm]{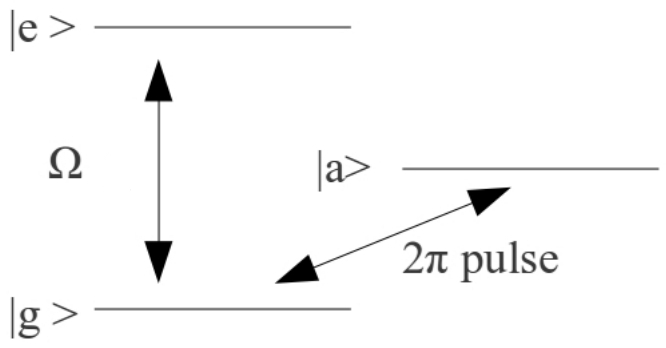}
\caption{System under study: two coupled levels and an auxilliary level}  
\end{figure}

\section{Large Detuning}
We examine a two level system, with levels labelled $\ket{g}$ and $\ket{e}$ coupled by a field which may or may not be on resonance, and real interaction parameter $\Omega$. We use a sequence of ideal 2$\pi$ pulses at times $0<\delta_1 T<\delta_2 T<\dots<\delta_i T<\dots<\delta_N T<T$ coupling $\ket{g}$ to an auxilliary level $\ket{a}$ in order to effect a $\pi$-phase change in $\ket{g}$ (Figures 1 and 2).  We will see that because of the coupling between the levels, this phase change leads to destructive interference between terms in the time evolution of the transition amplitude.  We examine the time evolution of the transition amplitude for a general pulse sequence characterized by $\delta_i$, and ask for the optimum sequence.

\begin{figure}
\centering 
\includegraphics[width=8.6cm]{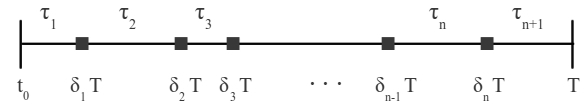}
\caption{Schematic diagram of pulse sequence.  Notice that $\delta_i$ quantifies the timing of the $i^{th}$ pulse and $\tau_i$ is the time interval between pulses $i$ and $i-1$.  That is, $\tau_i = T (\delta_i - \delta_{i-1})$.  Notice also that $T=\sum_{i=1}^{N+1} \tau_i$, where $\tau_i$ is a function of N such that $T$ is fixed.}
\end{figure}

	We use the semi-classical interaction Hamiltonian for a field of frequence $\omega'$ applied to a two-level system:
\begin{equation}
H(t) = \frac{\hbar \omega}{2} \sigma_z + \hbar \frac{\Omega}{4} e^{-i\omega't}\sigma^+ + \hbar \frac{\Omega}{4} e^{i\omega't}\sigma^- \label{hamiltonian}
\end{equation}
where $\sigma^+ \equiv ( \sigma_x +i\sigma_y)$, $\sigma^- \equiv ( \sigma_x - i\sigma_y)$, and $\Omega$ an interaction parameter, the Rabi frequency.  

In this section and Sec. IV, we will work in the interaction picture, because it is the interaction between the two-level system and the field we seek to suppress.  Of course, since the Schr\"odinger picture and interaction (or Dirac) picture are related by a unitary transformation, real observables or expectation values are unchanged.  Hence, we begin with Eqn.~\ref{hamiltonian} and write it as the sum of an unperturbed term, $H_0=\frac{\hbar \omega}{2} \sigma_z$, and a perturbation, or interaction term, $H_I(t)=\hbar \frac{\Omega}{4} e^{-i\omega't}\sigma^+ + \hbar \frac{\Omega}{4} e^{i\omega't}\sigma^-$.  Then the Hamiltonian becomes in the interaction picture:
\begin{align}
\tilde{H}_I(t) &= e^{i H_0 t/\hbar} H_I(t) e^{-i H_0 t/\hbar}\nonumber\\
&= \frac{\hbar\Omega}{4} \big[ \sigma^+ e^{i \Delta t} + \sigma^- e^{-i \Delta t} \big] \label{interaction hamiltonian}
\end{align}
where $\Delta \equiv \omega-\omega'$.

After transforming the state vector into the interaction picture using the transformation $\ket{\tilde{\Psi}}=e^{i H_0 t/\hbar}\ket{\Psi}$ and rewriting the Schr\"odinger equation in matrix form in terms of expansion coefficients in the interaction picture $\tilde{c_e}$ and $\tilde{c_g}$, we find the coupled first order differential equations:
\begin{align}
\dot{\tilde{c_g}}(t)&=-\frac{i\Omega}{2} e^{-i \Delta t} \tilde{c}_e(t)\nonumber\\
\dot{\tilde{c_e}}(t)&=-\frac{i\Omega}{2} e^{i \Delta t} \tilde{c}_g(t) \label{interaction diff eqns}
\end{align}

Under the approximation of large detuning, $\Delta>>\Omega$, we can use the machinery of first-order time-dependent perturbation theory.
\begin{align}
\tilde{c}_e(t)-\tilde{c}_e(t_0) &\equiv \Delta \tilde{c}_e (t,t_0)=- \frac{i \Omega}{2} \tilde{c}_g(t_0) \int_{t_0}^t e^{i \Delta t'} dt'  \nonumber\\
&= -\frac{\Omega}{2\Delta}(e^{i\Delta(t-t_0)}-1)e^{i\Delta t_0} \tilde{c}_g(t_0)
\end{align}

For the remainder of this section, we set $\tilde{c}_g(0)=1$, which corresponds to an atomic population intially in the ground state.  We can immediately reproduce the standard result for the probability of transition without a perturbing pulse sequence \cite{ge}:
\begin{equation}
|c_e(T)|^2=\Omega^2 \frac{\sin^2{(\Delta T/2)}}{\Delta^2}
\end{equation}

Note that we can see the mechanical effect of the $\pi$ phase shift on $\tilde{c}_g$; it sends $\Delta \tilde{c}_e (t,t_0) \rightarrow -\Delta \tilde{c}_e (t,t_0)$.  To introduce the effects of a pulse sequence, let $\tau_k=T(\delta_k-\delta_{k-1})$ be the $k^{th}$ time interval between the $k^{th}$ and $k-1^{th}$ pulses, and explicitly write out the evolution of the state vector.  For times before the first 2$\pi$ pulse is applied, the state vector evolves under the influence of the laser field as:
\begin{equation*}
t_0<t<t_0+\tau_1  \quad  \ket{\tilde{\Psi}(t)}=\ket{\tilde{g}}+\Delta \tilde{c}_e(t,t_0)\ket{\tilde{e}}
\end{equation*}
The first pulse is applied after $\tau_1$ has elapsed, effecting a $\pi$ phase change in the ground state,
\begin{equation*}
t=t_0+\tau_1  \quad  \ket{\tilde{\Psi}(t)}=-\ket{\tilde{g}}+\Delta \tilde{c}_e(t_0+\tau_1,t_0)\ket{\tilde{e}}
\end{equation*}
after which the system evolves under a sort of time reversal due to the phase change of $\ket{\tilde{g}}$, 
\begin{align*}
\text{for } t_0+\tau_1< &t <t_0+\tau_1+\tau_2 \\
\ket{\tilde{\Psi}(t)}=-\ket{\tilde{g}} + \Delta \tilde{c}_e(t_0 + &\tau_1,t_0)\ket{\tilde{e}} - \Delta \tilde{c}_e(t,t_0+\tau_1)\ket{\tilde{e}}
\end{align*}
until the next pulse is applied.
\begin{align*}
t= &t_0+\tau_1+\tau_2 \\
\ket{\tilde{\Psi}(t)} = &\ket{\tilde{g}}+\Delta \tilde{c}_e(t_0+\tau_1,t_0)\ket{\tilde{e}}\\
& -\Delta \tilde{c}_e(t_0+\tau_1+\tau_2,t_0+\tau_1)\ket{\tilde{e}}
\end{align*}
In general, after $N$-pulses:
\begin{align}
T = &t_0+\sum_{k=1}^{N+1} \tau_k  \nonumber\\
\ket{\tilde{\Psi}(T)} = &(-1)^N\ket{\tilde{g}}+ \nonumber\\
&\underbrace{\sum_{p=0}^{N} (-1)^{p} \: \Delta \tilde{c}_e \big(t_0+\sum_{k=1}^{p+1} \tau_k, \: t_0+\sum_{k=1}^p \tau_k \big)}_{\tilde{c}_e(t)}\ket{\tilde{e}}
\end{align}
Hence, upon substituting the explicit form of $\Delta \tilde{c}_e(t,t_0)$:
\begin{equation}
\tilde{c}_e(T) = -\frac{\Omega}{2\Delta} \sum_{p=0}^N (-1)^p (e^{i\Delta \tau_{p+1}}-1)e^{i \Delta \sum_{k=1}^p \tau_k} \label{transition probability tau}
\end{equation}
Upon making the change of variables $\tau_k=T(\delta_k-\delta_{k-1})$, Eqn.~\ref{transition probability tau} can be put into the form:
\begin{equation}
|c_e(T)|^2= \frac{\Omega}{2\Delta} \big[1 + (-1)^{N+1} e^{i \Delta T} + 2 \sum_{p=1}^{N} (-1)^p e^{i \Delta T \delta_p} \big] \label{optimize}
\end{equation}

One can show that $|c_e(T)|^2$ for an equidistant pulse sequence, $\delta_i=i/(N+1)$, is:
\begin{align}
|c_e(T)|^2 &= \frac{\Omega^2}{\Delta^2} \tan^2{\frac{\Delta T}{2N+2}}\cos^2{\frac{\Delta T}{2}} \: &\text{$N$ even}\\
|c_e(T)|^2 &= \frac{\Omega^2}{\Delta^2} \tan^2{\frac{\Delta T}{2N+2}}\sin^2{\frac{\Delta T}{2}} \: &\text{$N$ odd} \label{equidistant probability}
\end{align}
This corresponds to one of the key results of Agarwal et. al. \cite{ag}. One can see immediately that a large number of pulses suppresses the transition probability.  We now look for an optimum sequence $\delta_i$ such that the transition probability is minimized.  The expression inside the brackets of Eqn.~\ref{optimize} is identical to Eqn.~10 in Uhrig's paper~\cite{uh}, and the author uses the $N$ degrees of freedom afforded by $N$ pulses to set the first $N$-derivatives with respect to $T$ of Eqn.~\ref{optimize} equal to zero.  The result is a system of coupled equations which can be solved for $\delta_i$ to yield:
\begin{equation}
\delta_i = \sin^2 \bigg( \frac{\pi i}{2N+2} \bigg) \label{uhrig sequence}
\end{equation}

Plugging this expression for $\delta_i$ into Eqn.~\label{uhrig} gives an analytic expression for the probability of transition:
\begin{align}
|c_e(T)|^2 &= \bigg|\frac{\Omega}{2\Delta}\sum_{j=-n-1}^n (-1)^j e^{ (i\Delta T/2)\cos{(\pi j/(n+1)) } } \bigg|^2\\ \label{uhrig probability}
&\approx \frac{4 \Omega^2}{\Delta^2} (N+1)^2 J_{N+1}^2(\Delta T/2) 
\end{align}
The approximation follows from the Jacobi-Anger expansion, evaluation of a geometric series, evaluation of indeterminate terms as a limit, and finally approximating a series by its leading term.  Note that the approximation is good for $\Delta T < 2N+2$ \cite{uh}.  Figure 4 illustrates graphically the efficacy of the UDD compared to the equidistant sequence for a relatively small pulse number.

\section{Exact Solution}
We now solve the Rabi problem exactly in the rotating frame using the formalism of transfer matrices. Of course, since the Schr\"odinger picture and rotating frame are related by a unitary transformation, real observables or expectation values are unchanged.

To begin, we use properties of Pauli matrices to rewrite the Rabi Hamiltonian, Eqn.~\ref{hamiltonian}.
\begin{equation}
H(t) =\frac{\hbar \omega}{2} \sigma_z + \hbar \frac{\Omega}{2} (e^{-i\frac{\omega'}{2}t} \sigma_x e^{i\frac{\omega'}{2}t})
\end{equation}
To eliminate the time dependence of the Hamiltonian, we switch to a rotating frame.
\begin{equation}
\ket{\psi'} = e^{i\frac{\omega'}{2}t}\ket{\psi}
\end{equation}
Under this transformation the Schr\"odinger equation in the rotating frame becomes:
\begin{equation}
i \hbar \frac{\partial}{\partial t} \ket{\psi'} = \frac{\hbar \Delta}{2} \sigma_z \ket{\psi'} + \frac{\hbar \Omega}{2} \sigma_x \ket{\psi'}
\end{equation}
By writing our state vector in the two-level basis in the rotating frame, this becomes in matrix form:
\begin{equation}
\begin{pmatrix}
\dot{c}'_e\\
\dot{c}'_g
\end{pmatrix}
=\frac{i}{2}
\begin{pmatrix}
-\Delta & \Omega \\
\Omega & \Delta
\end{pmatrix}
\begin{pmatrix}
c'_e \\ 
c'_g
\end{pmatrix}
\label{rotating schrodinger}
\end{equation}

Eqn.~\ref{rotating schrodinger} is a system of coupled first order linear differential equations which can be solved to yield a generalized Rabi frequency, $\Omega_R \equiv \sqrt{\Delta^2+\Omega^2}$ and the time-evolution of the coefficients of expansion in the rotating frame. 
\begin{equation}
\begin{pmatrix}
c'_e(t)\\
c'_g(t)
\end{pmatrix}
=M(t)
\begin{pmatrix}
c'_e(0) \\ 
c'_g(0)
\end{pmatrix}
\end{equation}
where $M(t)$ is the unitary matrix:
\begin{equation*}
\begin{pmatrix}
\cos(\frac{1}{2} \Omega_R t) - \frac{i \Delta}{\Omega_R} \sin(\frac{1}{2}\Omega_R t) & \frac{i\Omega}{\Omega_R} \sin(\frac{1}{2}\Omega_R t) \\
\frac{i\Omega}{\Omega_R} \sin(\frac{1}{2}\Omega_R t) & \cos(\frac{1}{2} \Omega_R t) + \frac{i \Delta}{\Omega_R} \sin(\frac{1}{2}\Omega_R t)
\Omega
\end{pmatrix}
\end{equation*}

As a mathematical note, we can now see why the rotating frame was the best choice for solving the problem exactly. We use the rotating frame because the $SU_2$ group property has a desirable form when we transform to a picture in which the Hamiltonian is time independent; namely, for $M(\tau)$, $N(\tau)$ $\in SU_2$, $M(\tau_1)N(\tau_2)=MN(\tau_1+\tau_2)$.  This is desirable because in physical terms this says that the evolution for time $\tau_1$ followed by evolution for time $\tau_2$ is the same as evolution for total time $\tau_1+\tau_2$.

Now we define a matrix which represents the application of an ideal $2\pi$-pulse to the ground state through the auxilliary level.  Each of the two-levels is an eigenstate of the spin operator $\sigma_z$, so each state possesses the unique spinor property that a rotation by $2\pi$ produces a $\pi$ phase shift \cite{sl}.  Hence, a $2\pi$-pulse changes the sign of the ground state.
\begin{equation}
\Pi \equiv 
	\begin{pmatrix}
	1 & 0\\
	0 & -1
	\end{pmatrix}
= \sigma_z
\end{equation}
Using this formalism, we can easily write down an expression for $c'_g$ and $c'_e$ after an $N$-pulse sequence.
\begin{align}
	\begin{pmatrix}
	c'_e(T)\\
	c'_g(T)
	\end{pmatrix}
&= M(\tau_{N+1}) \, \Pi \, M(\tau_N)  \dots \Pi \, M(\tau_1) \,
	\begin{pmatrix}
	c'_e(0)\\
	c'_g(0)
	\end{pmatrix}\nonumber\\
&=  M(\tau_{N+1}) \, \prod_{i=1}^N \Pi M(\tau_i)
	\begin{pmatrix}
	c'_e(0)\\
	c'_g(0)
	\end{pmatrix}
\label{exact evolution}
\end{align}
In Sec. V we turn to the numerical study of Eqn.~\ref{exact evolution} for different pulse sequences, and hence different expressions $\tau_i$.  Numerical studies , such as Figure 5, suggest that the UDD is optimal over all ranges in detunings.

\section{Small Detuning}
We now examine the limit of zero-detuning, or equivalently, large pulse numbers..  If we say that $\tau \approx T/N$, this case is the limit $\frac{\Delta T}{N} \approx 0$.  We work in the interaction picture and solve the coupled differential equations in Eqn.~\ref{interaction diff eqns} after setting the product $\Delta t = 0$ and find, upon solving for coefficients in the interaction picture:
\begin{align}
\tilde{c}_g(t) &= \tilde{c}_g(t_0) \cos(\tfrac{1}{2}\Omega t) - i \tilde{c}_e(t_0) \sin(\tfrac{1}{2}\Omega t)\\
\tilde{c}_e(t) &= \tilde{c}_e(t_0) \cos(\tfrac{1}{2}\Omega t) - i \tilde{c}_g(t_0) \sin(\tfrac{1}{2}\Omega t) 
\end{align}

In the spirit of the previous section, we define another unitary transfer matrix $M(t)$:
\begin{equation}
M(t) \equiv
	\begin{pmatrix}
	\cos(\tfrac{1}{2}\Omega t)  & -i\sin(\tfrac{1}{2}\Omega t) \\
	-i\sin(\tfrac{1}{2}\Omega t) & \cos(\tfrac{1}{2}\Omega t) 
	\end{pmatrix}
\end{equation}

Hence, the time evolution of $\tilde{c}_g$ and $\tilde{c}_e$ between $t_0$ and $t$ is give by:
\begin{equation}
	\begin{pmatrix}
	\tilde{c}_e(t)\\
	\tilde{c}_g(t)
	\end{pmatrix}
=M(t)
	\begin{pmatrix}
	\tilde{c}_e(t_0)\\
	\tilde{c}_g(t_0)
	\end{pmatrix}
\end{equation}

Again, we describe pulses by matrices, $\Pi$, and we form an equivalent expression to Eqn.~\ref{exact evolution} in the interaction picture.
\begin{align}
	\begin{pmatrix}
	\tilde{c}_e(T)\\
	\tilde{c}_g(T)
	\end{pmatrix}
&=  M(\tau_{N+1}) \, \prod_{i=1}^N \Pi M(\tau_i)
	\begin{pmatrix}
	\tilde{c}_e(0)\\
	\tilde{c}_g(0)
	\end{pmatrix} \label{transfer eqn}
\end{align}

We can actually find a simple analytical solution to Eqn.~\ref{transfer eqn} quite easily.  The key is to simply starts multiplying out some of these matrices and use the trigonometric identities $ \sin{\alpha}\cos{\beta} \pm \sin{\beta}\cos{\alpha} = \sin{(\alpha \pm \beta)}$ and $\cos{\alpha}\cos{\beta} \mp \sin{\alpha}\cos{\beta} = \cos{(\alpha \pm \beta)}$.  Upon making the notational definition $\Theta_i \equiv \tau_i-\tau_{i-1}+\dots+(-1)^i\tau_1$ one can see by induction that Eqn.~\ref{transfer eqn} becomes:
\begin{align}
	\begin{pmatrix}
	\tilde{c}_e(T)\\
	\tilde{c}_g(T)
	\end{pmatrix}
&=	\begin{pmatrix}
	(-1)^N \cos(\tfrac{1}{2}\Omega \Theta_{N+1}) & -i\sin(\tfrac{1}{2}\Omega \Theta_{N+1})\\
	-i(-1)^N\sin(\tfrac{1}{2}\Omega \Theta_{N+1}) & \cos(\tfrac{1}{2}\Omega \Theta_{N+1})
	\end{pmatrix}
\nonumber\\ & \times
	\begin{pmatrix}
	\tilde{c}_e(0)\\
	\tilde{c}_g(0)
	\end{pmatrix}
\end{align}

Therefore, solving finally for the excited and ground state probabilities, we have:
\begin{align}
|c_g(T)|^2 &= |c_g(0)|^2 \cos^2(\tfrac{1}{2}\Omega \Theta_{N+1}) + |c_e(0)|^2 \sin^2(\tfrac{1}{2}\Omega \Theta_{N+1})\\
|c_e(T)|^2 &= |c_e(0)|^2 \cos^2(\tfrac{1}{2}\Omega \Theta_{N+1}) + |c_g(0)|^2 \sin^2(\tfrac{1}{2}\Omega \Theta_{N+1})
\end{align}

In order to decouple the levels from Rabi oscillations, we want to trap the state such that $|c_g(T)|^2 \approx |c_g(0)|^2$ and $|c_e(T)|^2 \approx |c_e(0)|^2$.  Hence, we seek a pulse sequence such that $\Theta_{N+1}$ is minimized.

Our expression for $\Theta_{N+1}$ can be put into a more useful form by the change of variables $\tau_i=T(\delta_i-\delta_{i-1})$.  We find that:
\begin{equation}
\Theta_{N+1} = T \big( 1+2(-1)^{N+1} \sum_{k=1}^N (-1)^k \delta_k \big) \label{small detuning optimize}
\end{equation}

To see how well the equidistant sequence does in decoupling the levels, we set $\delta_i=\frac{i}{N+1}$ and use the arithmetic identities:
\begin{align}
\sum_{k=1}^m (-1)^k k &= \frac{m}{2} \: &\text{m even}\\
\sum_{k=1}^m (-1)^k k &= -\frac{(m+1)}{2} \: &\text{m odd.}
\end{align}
We find that:
\begin{align}
\Theta_{N+1} &= \frac{T}{N+1} \: &\text{N even}\\
\Theta_{N+1} &= 0	      \: &\text{N odd}
\end{align}

Hence, we can decouple the levels to arbitrary precision given a large number of pulses.  In fact, any number of odd pulses will decouple the two levels completely in the sense that $|c_g(T)|^2 = |c_g(0)|^2$ and $|c_e(T)|^2 = |c_g(0)|^2$.  This is easy to understand in terms of the physical effects of the $\pi$ phase shift on the ground state. The pulses effects a sort of time reversal where the system retraces it's steps after the pulse is applied.  Hence, after an odd number of pulses, the system ends up exactly where it started.  Then why is the Uhrig pulse sequence optimum for a general qubit-bath system?  This is because the time reversal effected on the system affects only the qubit.  Hence, the qubit tries to retrace it's steps, but the bath is still constantly reacting to the instantaneous state of the qubit.  Since most of the time we assume that the initial qubit-bath state is a product state, after any finite time interval the system \emph{irreversibly} becomes a mixed state.  Therefore, the time reversal is imperfect, and the equidistant sequence is not optimum.  Then the question arises: what is the optimum sequence for any $N$ in our bathless system?

To answer this question, we seek a function $\delta_i$ such that:
\begin{equation}
(-1)^{N+1}+2 \sum_{k=1}^N (-1)^k \delta_k=0
\end{equation}

Which is simply setting Eqn.~\ref{small detuning optimize} equal to zero, dividing out nonzero $T$, and multiplying both sides by $(-1)^{N+1}$.  Upon inspection, one can see that the left hand side is actually the first derivative of Eqn.~\ref{optimize} at $\Delta T=0$.  By construction, the UDD satisfies the $k^{th}$ derivative equal to zero.  Specifically, the UDD satisfies the first derivative equal to zero.  Hence, the UDD again optimizes the decoupling of the levels in the sense that the system returns to it's initial state.

\begin{figure}[h]
\includegraphics[width=8.6cm]{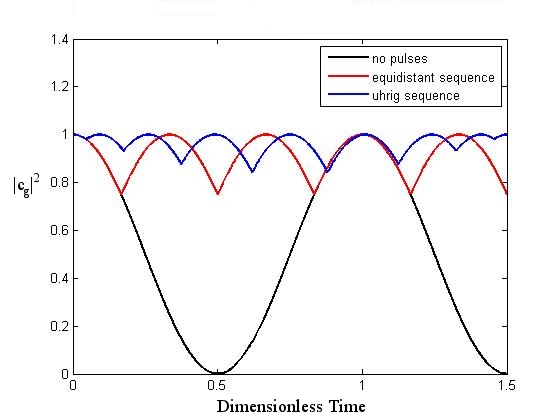} 
\caption{Plot of the evolution of the ground state probability $|c_g|^2$ versus time for 8 pulses and $\Delta/\Omega=0$.  Note that the Uhrig sequence is more effective than the equidistant sequence in preserving the intial state of the sytem at $t=1.5$ since $|c_g|^2=1$ only for the Uhrig sequence.}
\end{figure}

\section{Numerical Results}
We now examine the efficacy of different sequences for preserving the initial state of the system, that is, how close $|c_e(T)|^2 \approx |c_e(0)|^2$.  The pulse sequences to be studied are: (i) no pulses, (ii) equidistant sequence, $\delta_i= \frac{i}{N+1}$ and (iii) UDD sequence, $\delta_i=\sin^2(\frac{\pi i}{2N+2})$. The following plots illustrate the efficacy of each sequence for suppressing a transition to the excited state when the population is initially in the ground state.  Figure 3 shows Rabi oscillations as a function of time, and how they are affected by the pulse sequences.  The pulses effect a perfect time reversal on the system for zero detuning, and the UDD has a symmetry which always returns the system to where it started.  In contrast, the efficacy of the equidistant sequence depends on the parity of the pulse number.  Note that both sequences have some symmetry, which the equidistant sequence loses with non-zero detuning, causing the UDD to be more effective for even or odd pulse numbers.  Compare this figure to experimental data by Morton~\cite{mo}.  Figure 4 compares the evolution of the system for different pulse sequences for a large detuning and small pulse number.  Here the equidistant sequence offers a suppression of the transition by a factor of about $10^2$, and the UDD sequence suppresses the transition by a factor of $10^5$.  Note how the UDD causes the transition probability to remain relatively flat for a period of time; this is the result of the first $N$-derivatives being set to zero.

\begin{figure}[h!]
\centering
\includegraphics[width=8.6cm]{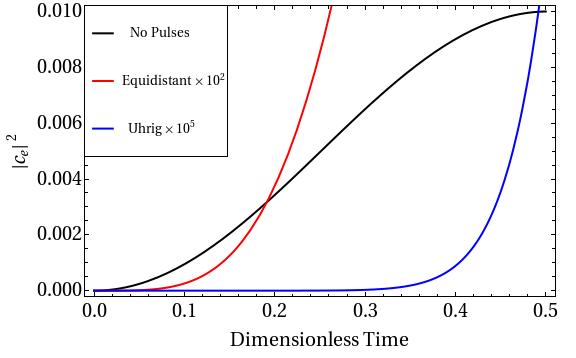}
\caption{For $n=5$ and $\Delta=10\Omega$, pictured is a comparison of transitions probabilities for no pulse sequence, equidistant sequence, and UDD sequence, respectively, as a function of time.}
\end{figure} 

Finally, Fig.5 compares the efficacy of the sequences over a range of pulse numbers, plotted with the analytic solutions for the large detuning case we derived.  It seems that the analytic solution for large detuning is still a good fit for this intermediate detuning.   The Uhrig sequence is more effective than the equidistant for suppressing the transition for all $n>1$.  For $n=1$, the UDD and equidistant sequences give the same result because the UDD degenerates to the equidistant sequence for $n=1$.  Note that each of the pulse sequences have been multiplied by an arbitrary factor of $20$ to make the graph more readable.  The transition probability with the UDD sequence drops very rapidly, reaching $\sim 10^{-2}$, $\sim 10^{-5}$,$\sim 10^{-10}$, and $\sim 10^{-14}$, for $n=2$, $4$, $6$, and $8$, respectively, compared to $\sim 10^{-2}$ for the equidistant sequence across this range of pulse numbers.  As a note, in Fig. 5 the system is allowed to evolve through half a Rabi cycle, that is $T=\pi/\Omega_R$, so the transition probability under no perturbing influence is a maximum for a given $\Delta/\Omega$.  In summary, the Uhrig pulse sequence proves to be many of orders of magnitude more effective for all values of $\Delta/\Omega$.

\begin{figure}
\includegraphics[width=8.6cm]{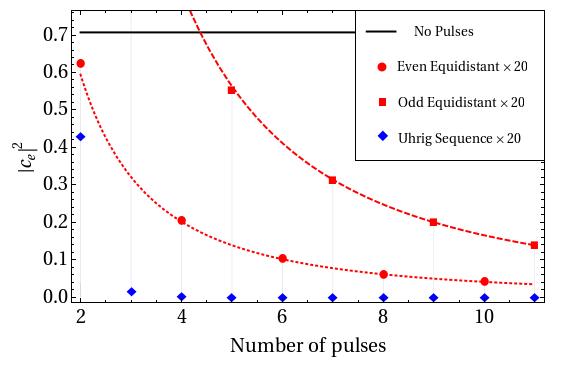}
\caption{Plot of the probability of transition $|c_e|^2$ versus pulse number for $\Delta/\Omega = 1$ for 2-11 pulses.}
\end{figure}

\vspace{15mm}
\section{Conclusion}

The decoupling of the levels of a two-level system has recently been realized in the experiments of Morton \cite{mo}.  In this experiment, nuclear fullerene qubits are decoupled from nuclear Rabi oscillations by taking advantage of the coupling between a nuclear and electronic qubit.  Through a repeated application of a very fast phase gate a nuclear spin qubit was bang-bang decoupled from a permanent driving field.  The speed of the phase gate and the resonance condition means that the equidistant sequence works very well in decoupling the system.  However, as we have shown, the UDD sequence offers an improvement over the equidistant sequence for any value of the detuning.  Hence, this decoupling scheme works well in tailoring the interactions in quantum systems, lending itself to broad applicability in quantum computing schemes \cite{qm}.  In addition to decoupling a system from its environment, the effect is required in some quantum computing schemes \cite{be}.

In conclusion, we have proved analytically that the Uhrig Dynamical Decoupling pulse sequence optimizes the inhibition of unwanted transitions in a two-level system for the case of large detuning as well as small detuning.  We have further confirmed our analytical results numerically. These results may prove fruitful for construction of a quantum computer, where suppression of decoherence and unwanted transitions are important considerations.  

\begin{acknowledgements}
C.A.S. thanks Oklahoma EPSCOR and Oklahoma State University for funding and support.
\end{acknowledgements}


\end{document}